\documentclass[12pt, twocolumn]{article}
\usepackage[utf8]{inputenc}
\usepackage[OT4]{fontenc}
\usepackage{amsmath}
\usepackage{amssymb}
\usepackage{graphicx}
\usepackage{multirow}
\usepackage{algorithmic}
\usepackage{subfigure}

\begin{document}
\title{Coarse-grained cellular automaton for traffic systems}
\author{Małgorzata J. Krawczyk and Krzysztof Kułakowski\\
Faculty of Physics and Applied Computer Science,\\
AGH University of Science and Technology,\\
al. Mickiewicza 30, 30-059 Cracow, Poland\\
gos@fatcat.ftj.agh.edu.pl, kulakowski@fis.agh.edu.pl}
\maketitle

\begin{abstract}
A coarse-grained cellular automaton is proposed to simulate traffic systems. There, cells represent road sections. A cell can be in two states: jammed or passable. 
Numerical calculations are performed for a piece of square lattice with open boundary conditions, for the same piece with some cells removed and for a map of a small city.
The results indicate the presence of a phase transition in the parameter space, between two macroscopic phases: passable and jammed. The results are
supplemented by exact calculations of the stationary probabilities of states for the related Kripke structure constructed for the traffic system. There,
the symmetry-based reduction of the state space allows to partially reduce the computational limitations of the numerical method.
\end{abstract}


\section{Introduction}

In simulations of traffic systems, cellular automata (CA) are a common tool \cite{rev1,rev2,rev4,rev3}. A cellular 
automaton consists a structure of cells, a set of cell states and the rule of time evolution which transfers a state
of a cell with its neighborhood to the state of this cell in the subsequent time moment. In this description, states,
space and time are discrete, while the traffic systems at lest space and time are inherently continuous. Still, there 
is a rich variety of CA which enable to investigate properties and phenomena of traffic systems, reproduced sometimes
with surprisingly subtle details. As a rule, the traffic CA are classified as single-cell or multi-cell models, 
where a vehicle occupies a single cell or more cells \cite{rev3}. On the other hand, traffic networks are also 
parametrized in different ways, as a node can represent a stop, a cross-road or a route \cite{spaces}. So simplified 
when compared with real systems, the technology of CA suffers known computational limitations: a more detailed 
description is paid by the smaller size of a simulated system.\\

More recently, a modification of CA has been applied where a cell represents a state of the whole considered system 
\cite{cpl05,gao}. This approach can be seen as an example of the concept of Kripke structures \cite{kri}, where nodes 
represent states of the whole system and links represent processes leading from one state to another. The obvious 
drawback of this parametrization is that the number of nodes increases exponentially with the system size. In 
\cite{cpl05}, this difficulty is evaded by taking into account only the states which appear during the time evolution. 
The same idea was developed into a technique of time series analysis, known as recurrence networks. Briefly, the rule
of time evolution is used to generate new states which are attached as nodes to the simulated network; in this way
the signal is characterized in terms of a growing network. For details see \cite{recnets} and references cited 
therein.\\

Our aim here is to discuss a new cellular automaton designed for modeling jams in traffic systems. The novelty of this 
automaton is that cells represent sections of road which can be either jammed or passable. A jam can grow at its end 
and flush at its front; the competition between these two processes depends on the local topology of the traffic network. 
Our description, inspired by percolation, is more coarse-grained, than in other models. According to the classification 
of traffic models, presented in \cite{www}, our model belongs to macroscopic queueing models. Some model elements 
remind the cell transmission model by Daganzo \cite{dag1,dag2}: namely, the rates of inflow and outflow in the cell transmission
model are similar to the rates of grow and flush of traffic jam, defined below. However, as it is explained in details 
in the next section, it is only jammed and passable cells what is differentiated here, and the flows of vehicles are not identified.
The price paid is that a range of dynamic phenomena as synchronization and density waves are excluded from the modeling. 
These phenomena, essential at the scale of a road \cite{e1,e2,e3,e4}, can be less important at the scale of a city. 
Consequently, our approach should be suitable for macroscopic modeling of large traffic systems.  \\

Our second aim is to construct the Kripke structure on the basis of the same cellular automaton. This in turn limits 
again the size of the system, because of the exponential size dependence of the number of states. We are going to
demonstrate that our recent tool, i.e. the symmetry-induced reduction of the network of states \cite{m1,m2}, is 
useful to partially reduce the computational barrier.\\

In the next section we describe the automaton in general terms and we recapitulate the method of reduction of the network of states, mentioned above.
As the direct form of the automaton rules depends on the traffic system under considerations, the exact description of the rules is 
given in Section 3, together with the information on the analyzed systems, both artificial (the square lattice) and real (a small city).
Two last sections are devoted to the numerical results and their discussion.

\section{The model}

\subsection{Cellular automaton}

We analyze a simple automaton, where each cell $i$- a road section - can be either in the state $s=0$ or $s=1$. The state 
$s_i=0$ means that a fluent motion via a given road section is possible, while the state 
$s=1$ means a traffic jam. As each road section is a part of larger system, the cell state depends 
on the state of roads where one can enter from a given road section. Namely, the probability of a traffic jam back propagation as well 
as the probability of a traffic jam to be flushed depend on the number and state of neighboring road sections.  To initialize calculations, 
one has to assign values of three parameters. Two of them, $w$ and $v$, describe the whole system, and the last one $p$ is related to 
the boundaries. Specifically, $w$ is the probability that a traffic jam arises on a given road section
due to its presence on the roads directly preceding the currently considered one (jam behind jam), $v$ is the probability of a jam flush 
(jam behind passable gets passable), and $p$ is the probability 
that a traffic jam appears at a road section at the boundary, but out of the system. The latter parameter describes the system interaction 
with the outer world. The parameters $w$ and $v$ can be related to the flows
used in \cite{h1,h2} for the discussion for congestion near on-ramps.\\

The probability of a change of the state of a given road section is obtained as the result of the analysis of the state of this section
and the state of its neighborhood. We ask for which ranges of the parameters the system is passable in the stationary state. \\

The detailed realization of the model depends of the topology of the traffic network. In Section 3 the exact algorithm is presented 
together with the presentation of the analyzed systems.

\subsection{The state space and its reduction}

The automaton defined above can be used for simulations, and the results of these simulations are reported below. The same automaton is used here
also to construct the network of states, as in \cite{cpl05,m1,m2}. This network, equivalent to the Kripke structure \cite{kri}, is formed by all 
possible combinations of states of roads which play 
the role of nodes. Next, an appropriate master equation \cite{vkamp} is constructed, which reflects all possibilities of states obtained from the 
current state.  The obtained matrix of transitions between states, i.e. the transfer matrix, is equivalent to the connectivity matrix of our state network.
For a given set of model parameters 
($p$, $w$ and $v$), eigenvector of the matrix associated with the eigenvalue equal 1 serves to calculate probabilities of particular states in the stationary state. 
Having these values, one can evaluate how passable the system is under given conditions from the average number of unjammed ($s=0$) states 
\begin{equation}
P(0)=\dfrac{1}{N}\sum\limits_{i=1}^{2^N}P_in_i(0)
\label{e1}
\end{equation}
where: $N$ is the size of the system (the number of considered road sections), $P_i$ - probability of $i$-th state and $n_i(0)$ - number of zeros in  
$i$-th state. We note that in this equation, me make an average over the states of the whole network, and not over the states of local cells.\\

As the obtained number of states is large even for moderate systems, we reduce the system size by the application of the procedure proposed in our previous 
papers \cite{m1,m2}. The method of the reduction of the system size is based on the symmetry observed in the system, which manifests in the fact 
that properties of some elements of the system are exactly the same. The starting point is the network of states, and the core of the method is  
to divide nodes into classes; the stationary probability of each node in the same class is the same \cite{m1,m2}. To begin, for each node  
the list of its neighboring nodes is specified, with the consideration of weights of particular connections. Provisionally, the class of each state 
is determined by its degree; for each state its symbol is replaced by the symbol of class, which discriminate nodes which have different number 
of neighbors. At the next stage we examine the lists of neighboring nodes in terms of class symbols assigned to a particular neighbors and weights 
of appropriate ties. If for nodes assigned with the same symbol the symbols assigned to its neighbors are different or their are the same but their weights 
are different, an additional class distinction is introduced. At the end of the algorithm, the classes, i.e. subsets of nodes are indicated, 
which have identical lists of neighbors with respect of the number of neighbors, the symbol assigned to each of them, and weights of particular 
connections \cite{m1,m2}.

\section{Analyzed systems}
\subsection{Square lattice}
As a reference system we analyze a system of directed roads placed on edges of a regular square lattice. The lattice is finite, with open boundary conditions. 
For such a system each road has two in-neighbors 
and two out-neighbors. As it will be explained in detail, the probability of the state change depends only on the state of out-neighbors. 
At the boundaries, a road has one or none out-neighbors (none at the corner). At each road, the traffic takes place only in one direction, 
say upwards and right. This setup is borrowed from the Biham-Middleton-Levine automaton \cite{bml}. The algorithm of a change of the state of the road for the square 
lattice is presented in Fig.\ref{alg1}.
\begin{figure}
{\footnotesize{
\begin{algorithmic}
\IF{$s=0$}
  \IF{$L_n=0$}
    \STATE $0\xrightarrow{P=pw}1$
  \ELSIF{$L_n=1$}
    \IF{$s_n=0$}
      \STATE $0\xrightarrow{P=p\dfrac{w}{2}}1$
    \ELSE
      \STATE $0\xrightarrow{P=p\dfrac{w}{2}+\dfrac{w}{2}}1$
    \ENDIF
  \ELSE
    \IF{$s_{n1}\ne s_{n2}$}
      \STATE $0\xrightarrow{P=\dfrac{w}{2}}1$
    \ELSIF{$s_{n1}=s_{n2}=1$}
      \STATE $0\xrightarrow{P=w}1$
    \ENDIF
  \ENDIF
\ELSE
  \IF{$L_n=0$}
    \STATE $1\xrightarrow{P=v(1-p)}0$
  \ELSIF{$L_n=1$}
    \IF{$s_n=0$}
      \STATE $1\xrightarrow{P=\dfrac{v}{2}(2-p)}0$
    \ELSE
      \STATE $1\xrightarrow{P=\dfrac{v}{2}(1-p)}0$
    \ENDIF
  \ELSE
    \IF{$s_{n1}\ne s_{n2}$}
      \STATE $1\xrightarrow{P=v}0$
    \ELSIF{$s_{n1}=s_{n2}=1$}
      \STATE $1\xrightarrow{P=\dfrac{v}{2}}0$
    \ENDIF
  \ENDIF
\ENDIF
\end{algorithmic}
}}
\caption{Algorithm of a change of the state of the road for the square lattice.}
\label{alg1}
\end{figure}

In the above algorithm $s$ is a state of a given road, $L_n$ is the number of roads given road is connected to, the quantity $s_n$ refers to the state of the road 
which is a neighbor of the currently considered one (if the road has two neighbors their states are marked respectively as $s_{n1}$ and  $s_{n2}$). The probability 
$P$ of a change of the state depends, as it was mentioned above, on the state of the considered road and the state of the neighborhood which determine how probable 
the change is. Namely, the transition from jammed to passable (1 to 0) mean that the jam at a given road section is flushed by free motion of vehicles at the jam front.
This is possible only if the out-neighboring section is empty. Further, the transition from passable to jammed (0 to 1) is possible only if the out-neighboring section 
is jammed. In both cases, the transition depends on the state of the out-neighbors; the state of in-neighbors is not relevant.\\

\begin{center}
\begin{figure}
\includegraphics[width=.4\columnwidth, angle=0]{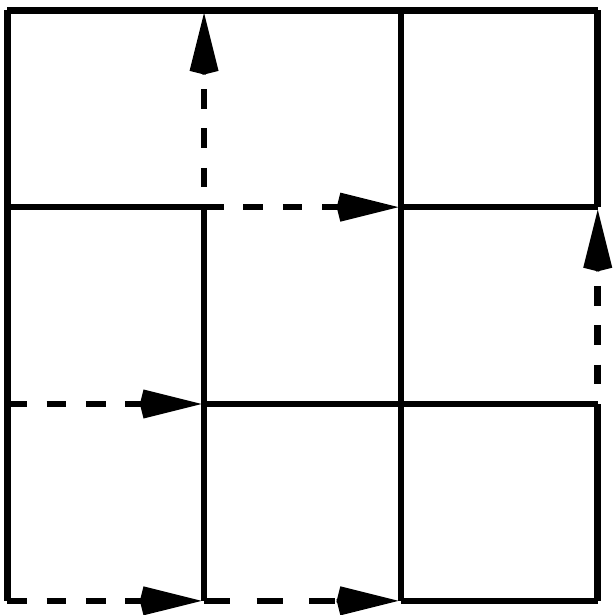}
\begin{minipage}{.1\columnwidth}
\vspace{-3cm}
\centering$\Rightarrow$
\end{minipage}
\includegraphics[width=.4\columnwidth, angle=0]{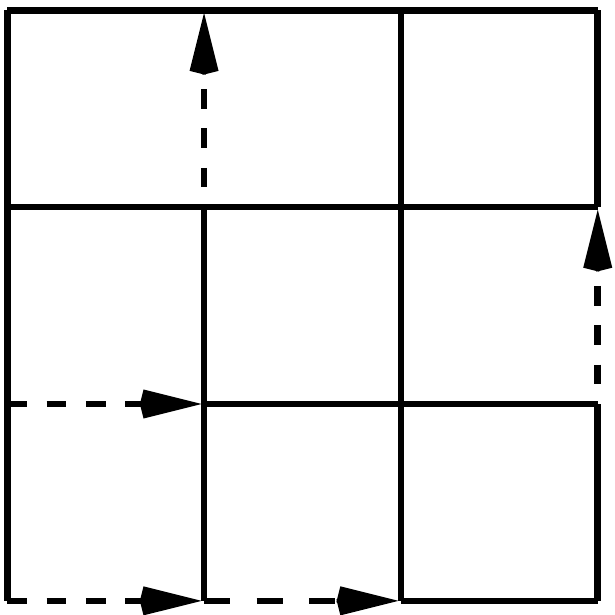}
\caption{Example of the change of traffic. A dashed-line refers to the state $0$ (fluent flow), and a solid-line refers to the state $1$ (traffic jam). Arrows 
indicate the direction of traffic.}
\label{art}
\end{figure}
\end{center}

In Fig.\ref{art} a piece of the system is presented. Each road can be either passable (in our notation a road is in a state $0$) which is marked as 
a dashed-line or a traffic jam can be formed (a road is in a state $1$) which is marked as a solid-line. The direction of traffic was ticked on roads in the state $0$, 
but the rule is the same for all roads; we keep down-up and left-right direction. Here we present one of the possible changes of the state of the system.

\subsection{Small city}

\begin{figure}[!hptb]
\begin{center}
\includegraphics[width=.4\textwidth, angle=0]{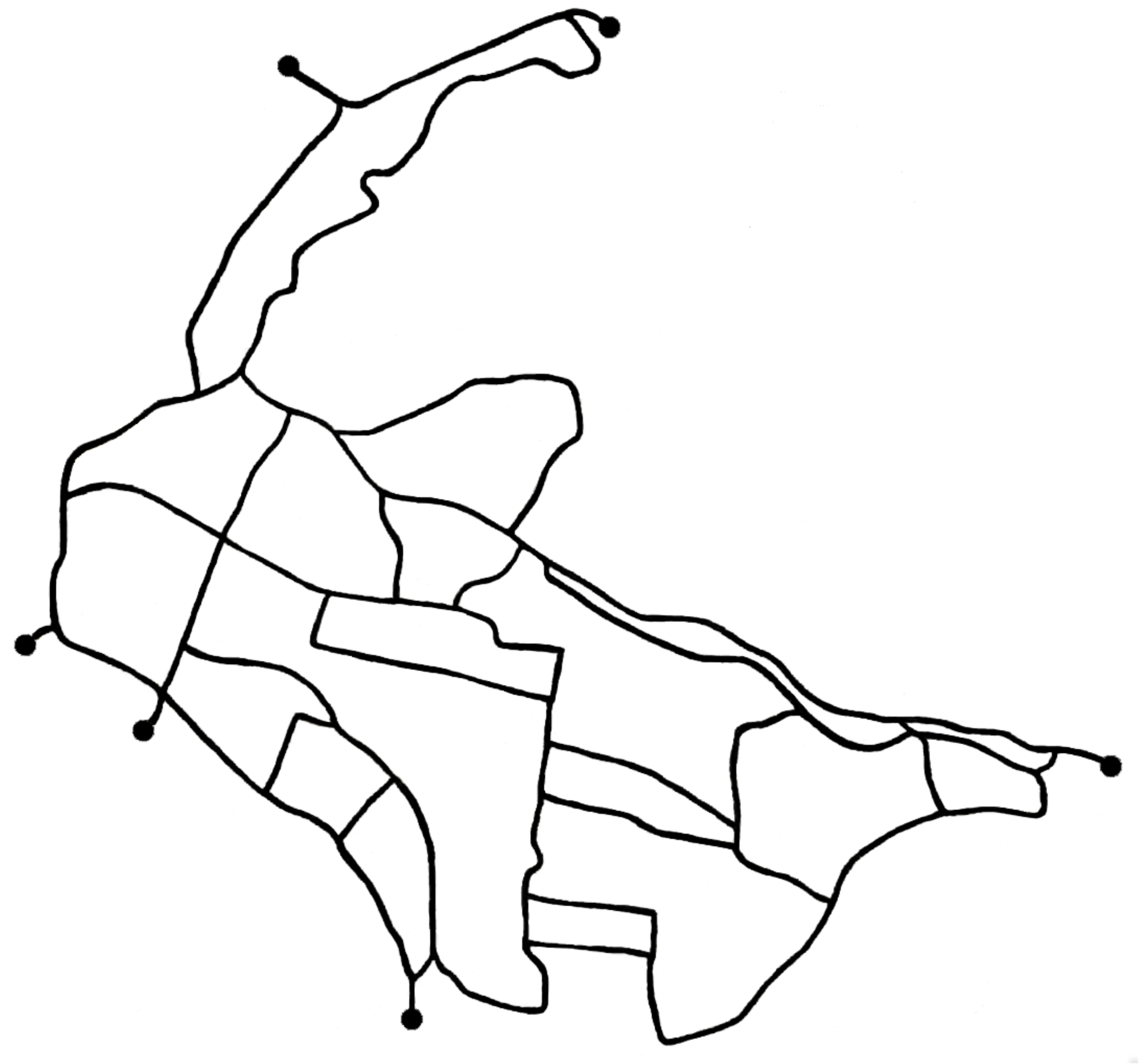}
\caption{Road network of Rabka. Roads are divided into $374$ approximately equal sections. The black dots mark the exit/entrance roads.}
\label{map}
\end{center}
\end{figure}

The method was also applied to a real road network of a small Polish town Rabka. The structure of roads which matter in traffic was selected - dead ends are removed (Fig.\ref{map}).
Each road, if necessary, was divided into sections of approximately equal length. We get $374$ sections. Here the number of neighbors for different roads varies as it 
results from the town topology. Each section is a two-way street. In this case the algorithm has a form presented in Fig.\ref{alg2}.

\begin{figure}
{\footnotesize{
\begin{algorithmic}
\IF{$s=0$}
  \IF{$L_n=0$}
    \STATE $0\xrightarrow{P=p\dfrac{w}{2}}1$
  \ELSE
      \STATE $0\xrightarrow{P=\dfrac{\sum\limits_{out}s}{k_{out}}\dfrac{w}{2}}1$
  \ENDIF
\ELSE
  \IF{$L_n=0$}
    \STATE $1\xrightarrow{P=v(1-p)}0$
  \ELSE
      \STATE $1\xrightarrow{P=v\sum\limits_{out}(1-s)}0$
  \ENDIF
\ENDIF
\end{algorithmic}
}}
\caption{Algorithm of a change of the state of the road for a real network.}
\label{alg2}
\end{figure}

In the algorithm presented in Fig.\ref{alg2} summing goes through the states of the roads outgoing from a given one, and $k_{out}$ is a number of outgoing roads.

\section{Results}
\label{results}
\subsection{The square lattice}

\begin{figure}[!hptb]
\begin{center}
\subfigure[$p=0.1$]{
\includegraphics[width=.5\textwidth, angle=0]{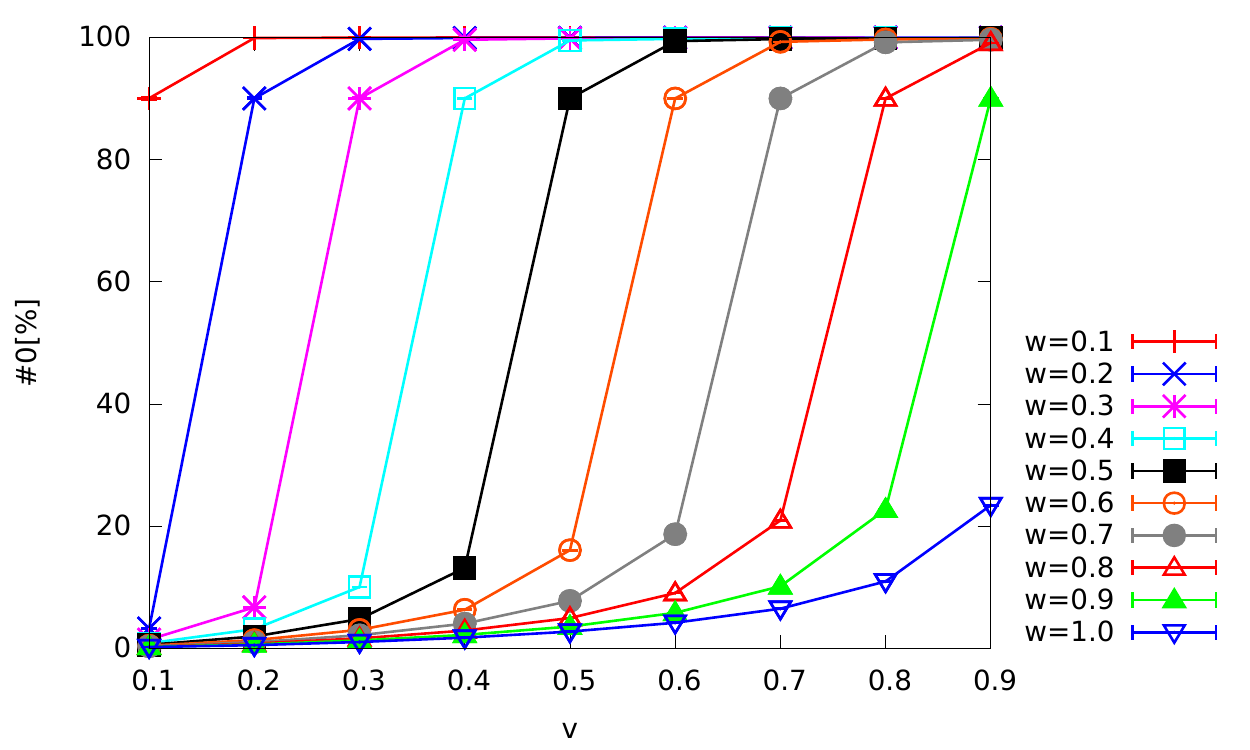}
\label{fig:3A}
}
\subfigure[$p=0.7$]{
\includegraphics[width=.5\textwidth, angle=0]{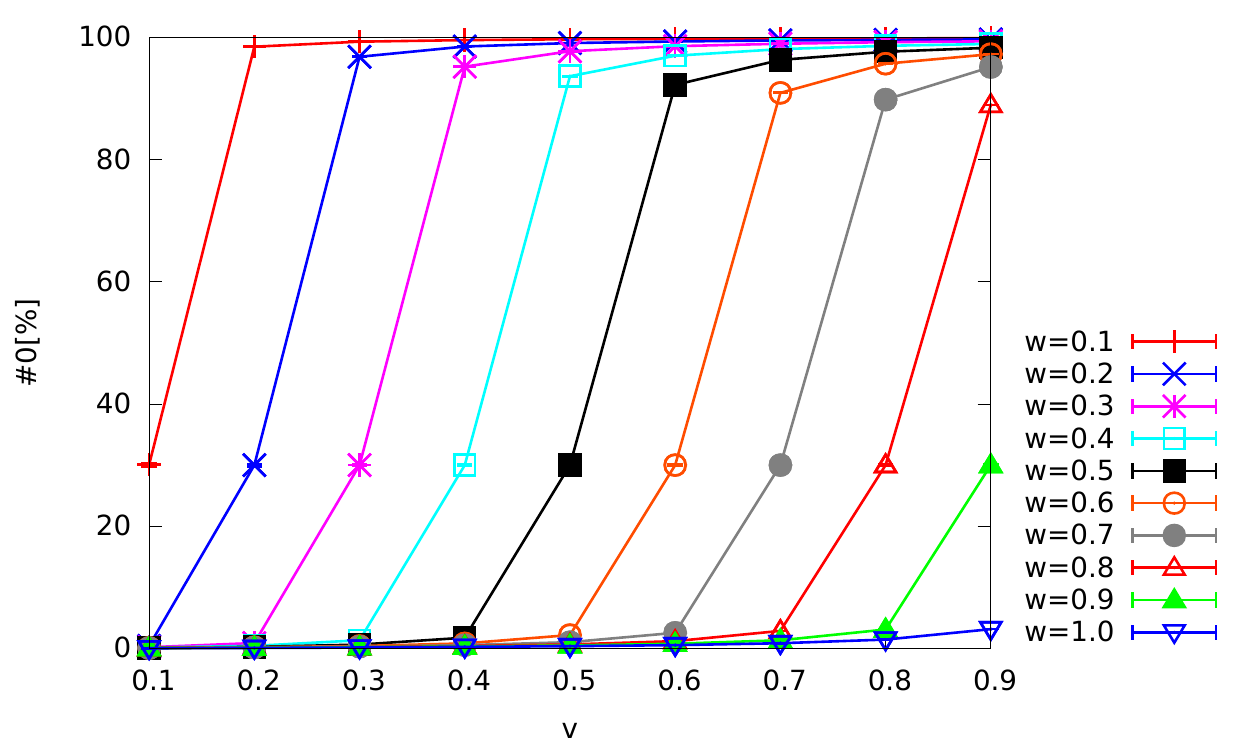}
\label{fig:3B}
}
\caption{Diagram for the square lattice of the size $100\times100$ (average in the steady state over $100$ realizations).}
\label{sq}
\end{center}
\end{figure}

All presented results are a time average in the steady state over $100$ realizations for the square lattice of the size $N=100\times100$. To check that 
the results do not depend on the initial conditions, we use three options for the initial state: states of all roads set to $0$, states of all roads set to $1$, 
and a state of each road is set randomly to be $0$ or $1$.\\
The results depend on the values of the model parameters $p$, $w$ and $v$. As a result for the whole system the percentage of roads in the state $0$ ($\#0[\%]$) 
is calculated. The higher the number of zeros the more passable the system is. The results for two different values of the parameter $p$ are presented in Fig. 
\ref{sq} for $p=0.1$ and for $p=0.7$.

\begin{figure}[!hptb]
\begin{center}
\includegraphics[width=.4\textwidth, angle=0]{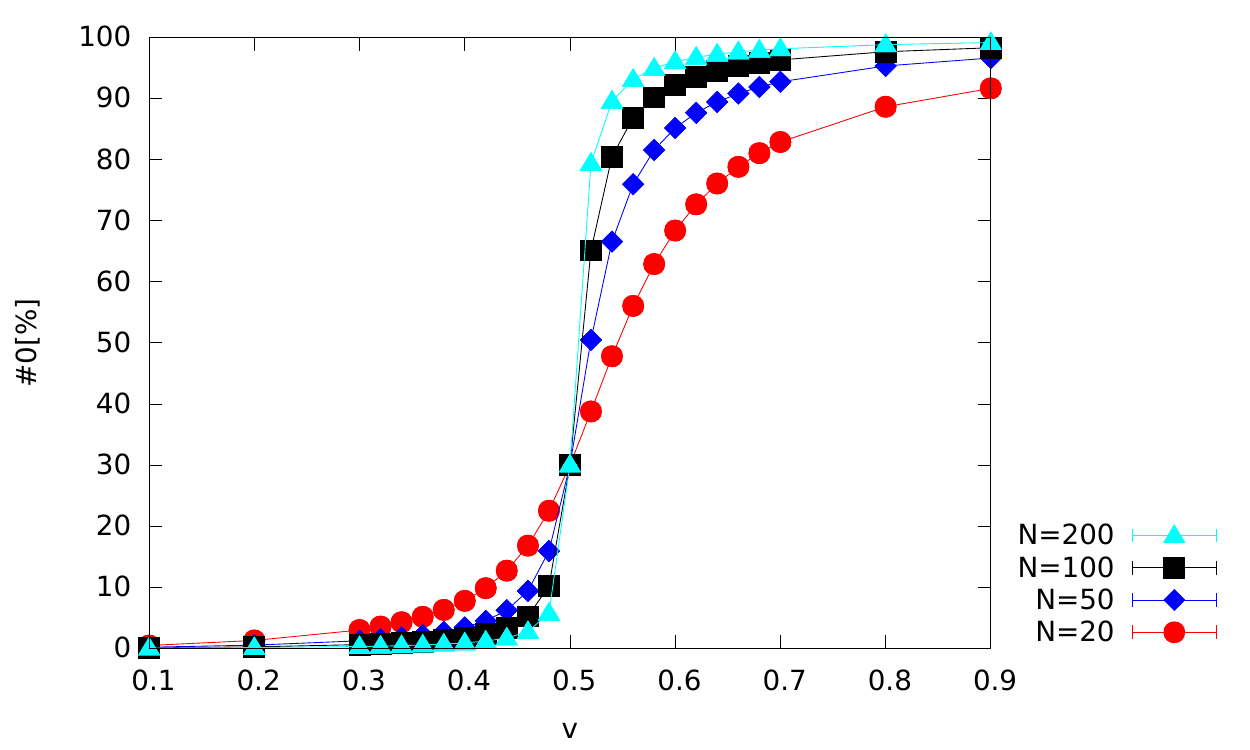}
\caption{Percentage of zeros in a function of $v$ for square lattices of different sizes $N\times N$ for $p=0.7$ and $w=0.5$ (average in the steady state over $100$ realisations).}
\label{pf}
\end{center}
\end{figure}

The increase of the percentage of zeros with the parameter $v$, visible in Fig.\ref{sq}, can be interpreted as an indication of a phase transition. To verify its sharpness 
dependence on the system size, we calculated the curve $\#0$ vs $v$ for a selected case: $p=0.7$, $w$=0.5 and various system sizes $N^2$. The results are shown in Fig.\ref{pf}. Indeed,
the sharpness increases with $N$, and the curve for $N=200$ is close to the step function.\\

\begin{figure}[!hptb]
\begin{center}
\subfigure[$p=0.1$]{
\includegraphics[width=.5\textwidth, angle=0]{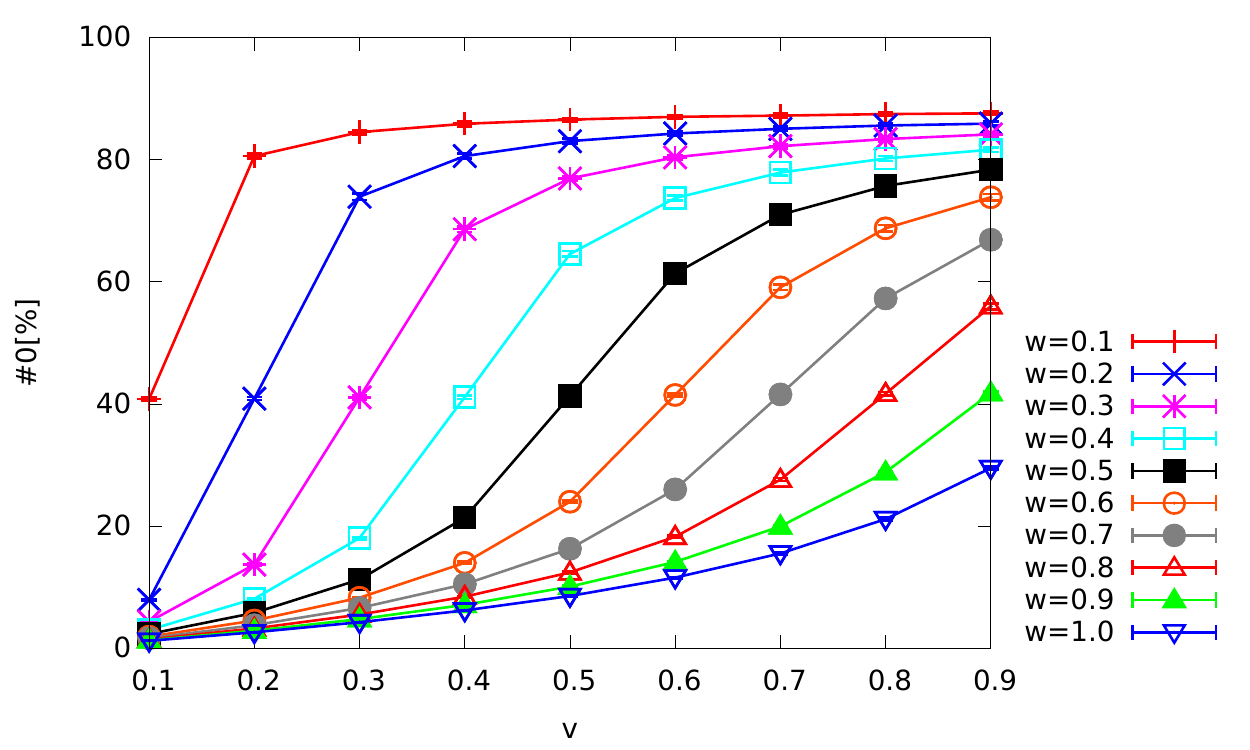}
\label{fig:4A}
}
\subfigure[$p=0.7$]{
\includegraphics[width=.5\textwidth, angle=0]{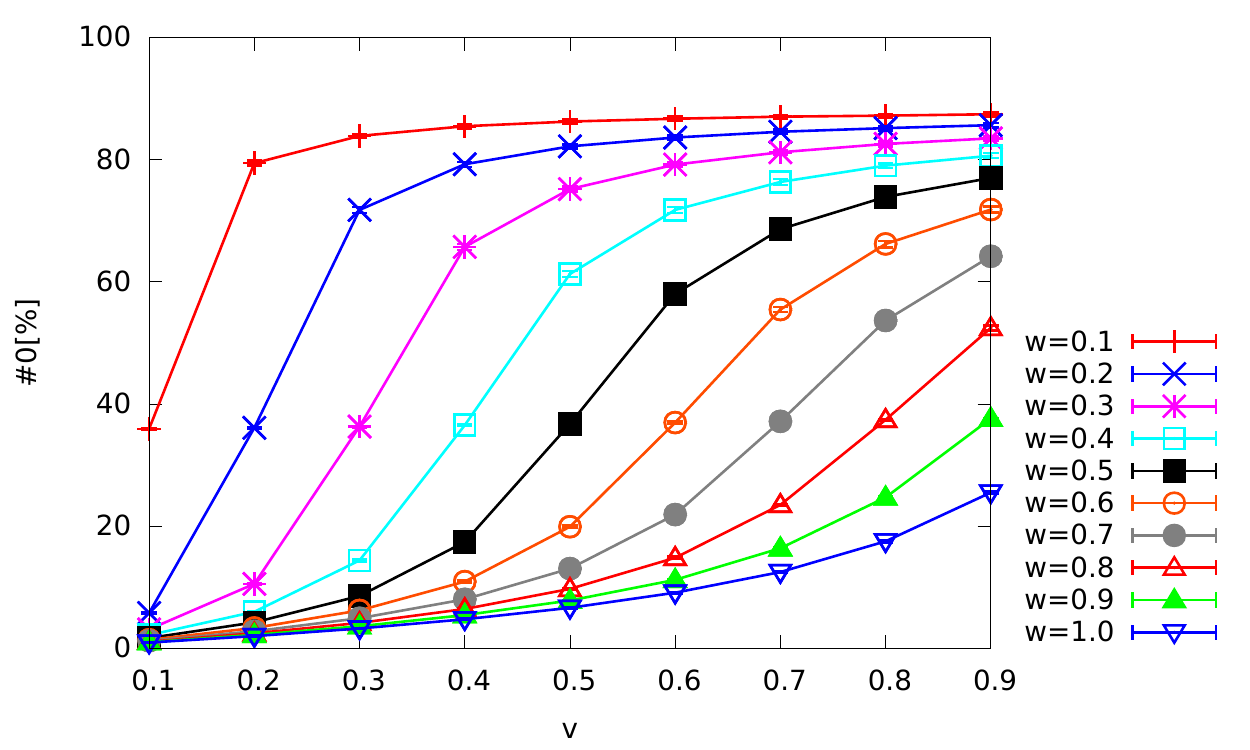}
\label{fig:4B}
}
\caption{Diagram for the square lattice of the size $100\times100$ (average in the steady state over $100$ realisations), with removal of $10\%$ 
randomly chosen road sections.}
\label{fig4}
\end{center}
\end{figure}

We also check how removal of some number of roads changes the obtained results, to check if the symmetry of the square lattice is necessary. In Fig.\ref{fig4} the results obtained 
when $10\%$ randomly chosen road sections is removed.
The removal is done separately for each realization. If, in a consequence of the removal, some part of lattice is isolated, it is removed as well. The results, shown in Fig.\ref{fig4},
indicate that the phase transition, found for the square lattice, is observed also in a randomized lattice. The maximal number of zeros in this case is less than 90 percent, because 
the plot is normalized to the whole square lattice, including the removed links.

\subsection{Small city}
\begin{figure}[!hptb]
\begin{center}
\subfigure[$p=0.1$]{
\includegraphics[width=.5\textwidth, angle=0]{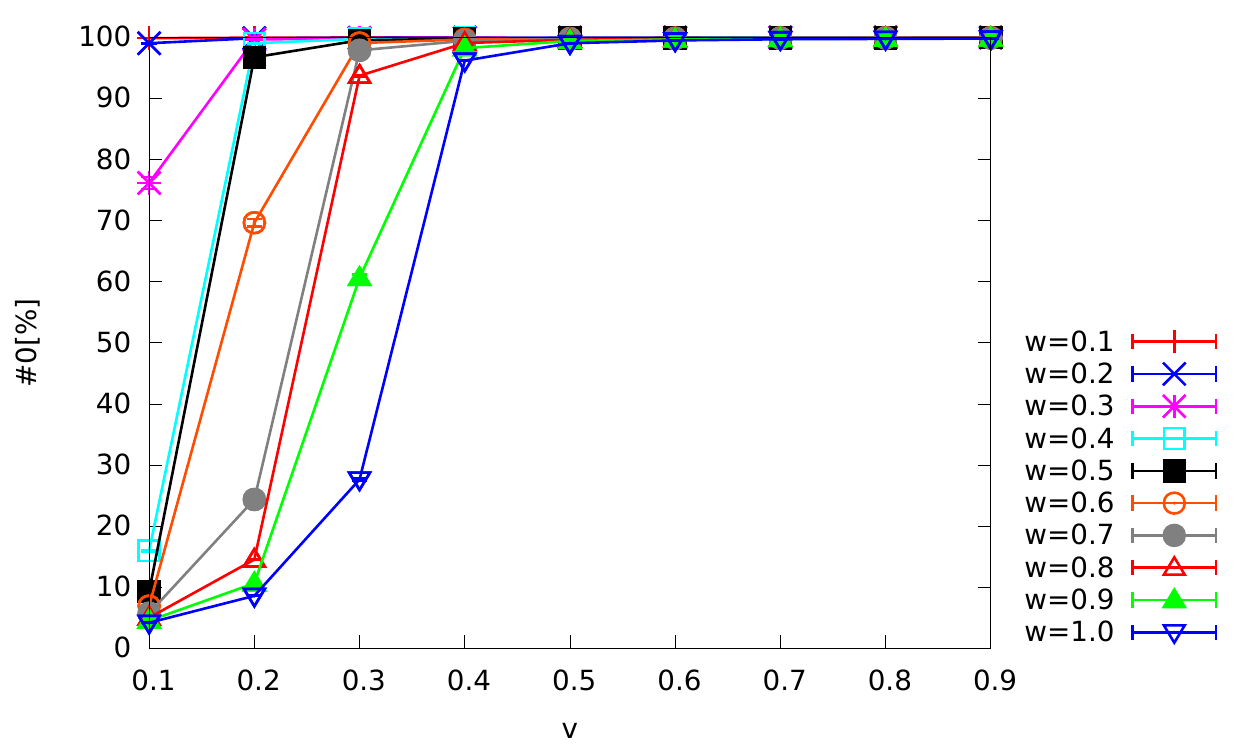}
\label{fig:5A}
}
\subfigure[$p=0.7$]{
\includegraphics[width=.5\textwidth, angle=0]{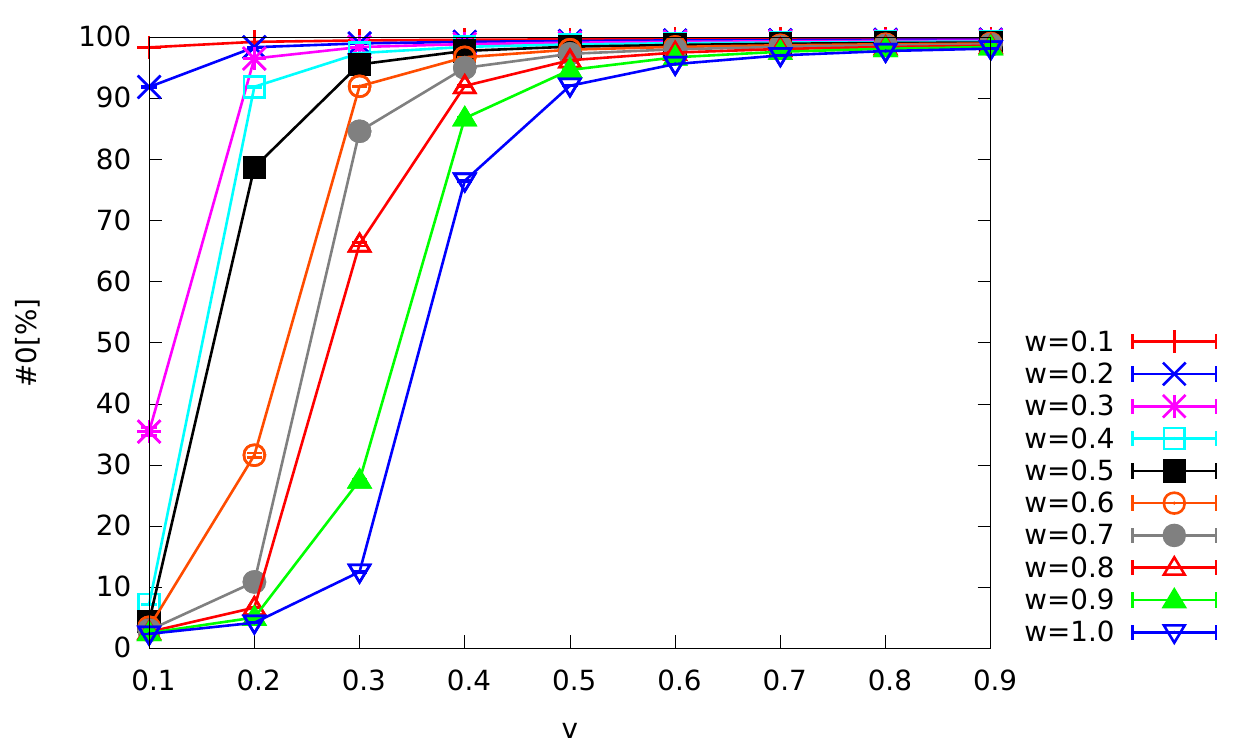}
\label{fig:5B}
}
\caption{Diagram for the real network of the size $N=374$.}
\label{fig5}
\end{center}
\end{figure}

The result obtained for the simulations of the traffic network in Rabka, formed by $374$ road sections, are presented in Figs.\ref{fig5}. The main difference
between this network, as constructed from the map in Fig.\ref{map}, and the square lattice (with removals or not) is that the Rabka network is less connected.
There, often the road sections form long chains. Comparing Figs.\ref{sq} and \ref{fig4} we see that the consequence of this difference is that jammed state is 
less likely. The origin of this result is that jams are created behind the jammed road sections; the more in-neighbors of these sections, the more jams appear. 
Besides of that, the obtained plot are similar to those for the square lattice.\\

\subsection{A simplified map}
\begin{figure}[!hptb]
\begin{center}
\subfigure[$p=0.1$]{
\includegraphics[width=.5\textwidth, angle=0]{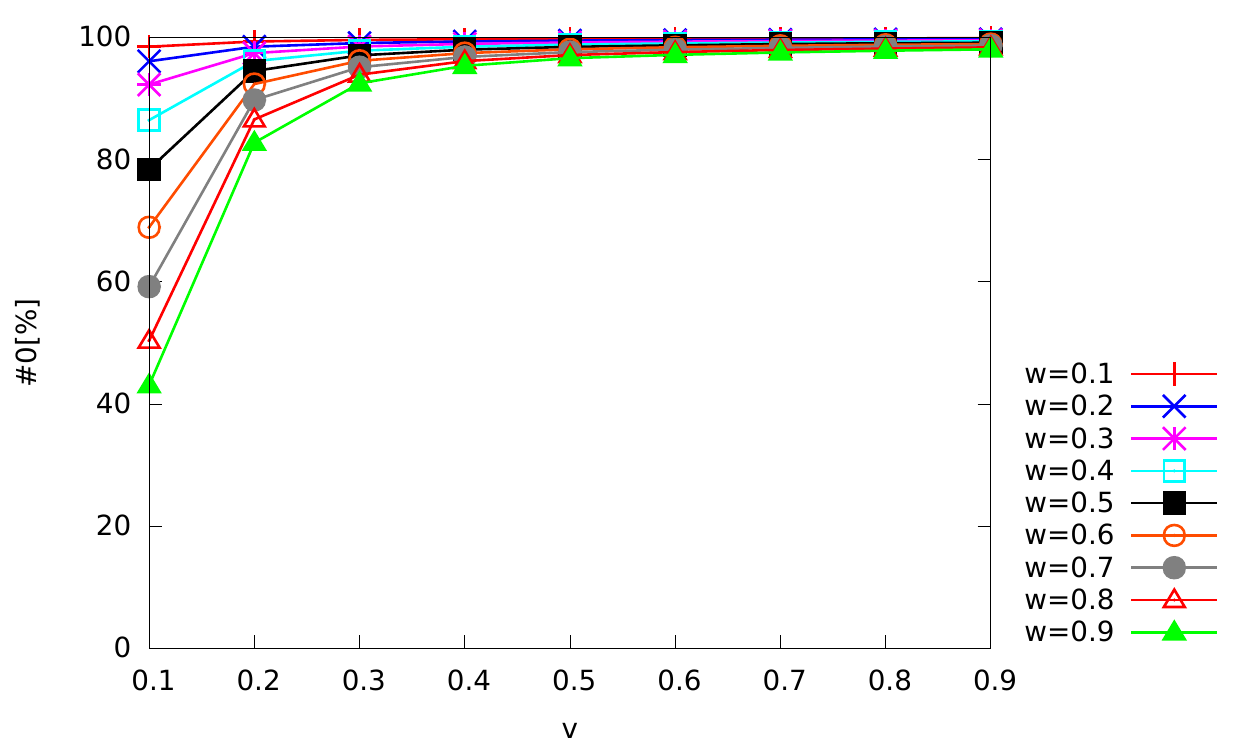}
\label{fig:6A}
}
\subfigure[$p=0.7$]{
\includegraphics[width=.5\textwidth, angle=0]{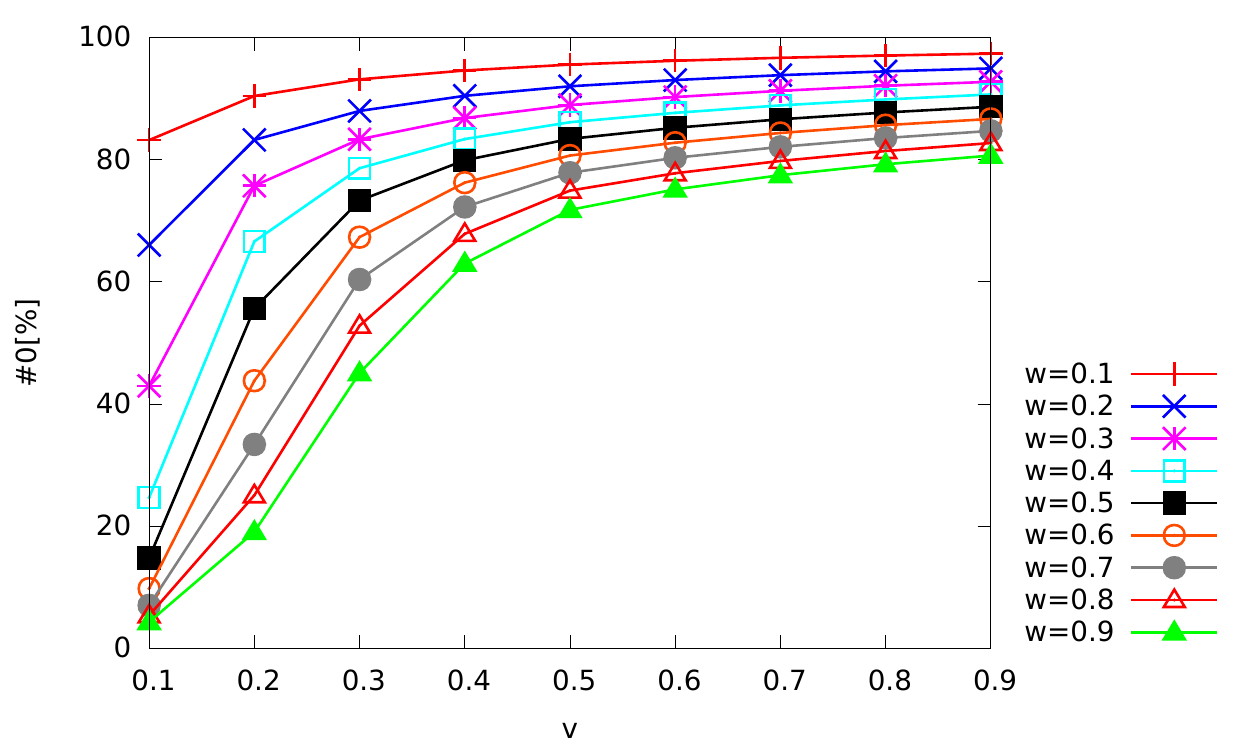}
\label{fig:6B}
}
\caption{Diagram for the real reduced network of the size $N=18$.}
\label{fig6}
\end{center}
\end{figure}

\begin{figure}[!hptb]
\begin{center}
\subfigure[$p=0.1$]{
\includegraphics[width=.5\textwidth, angle=0]{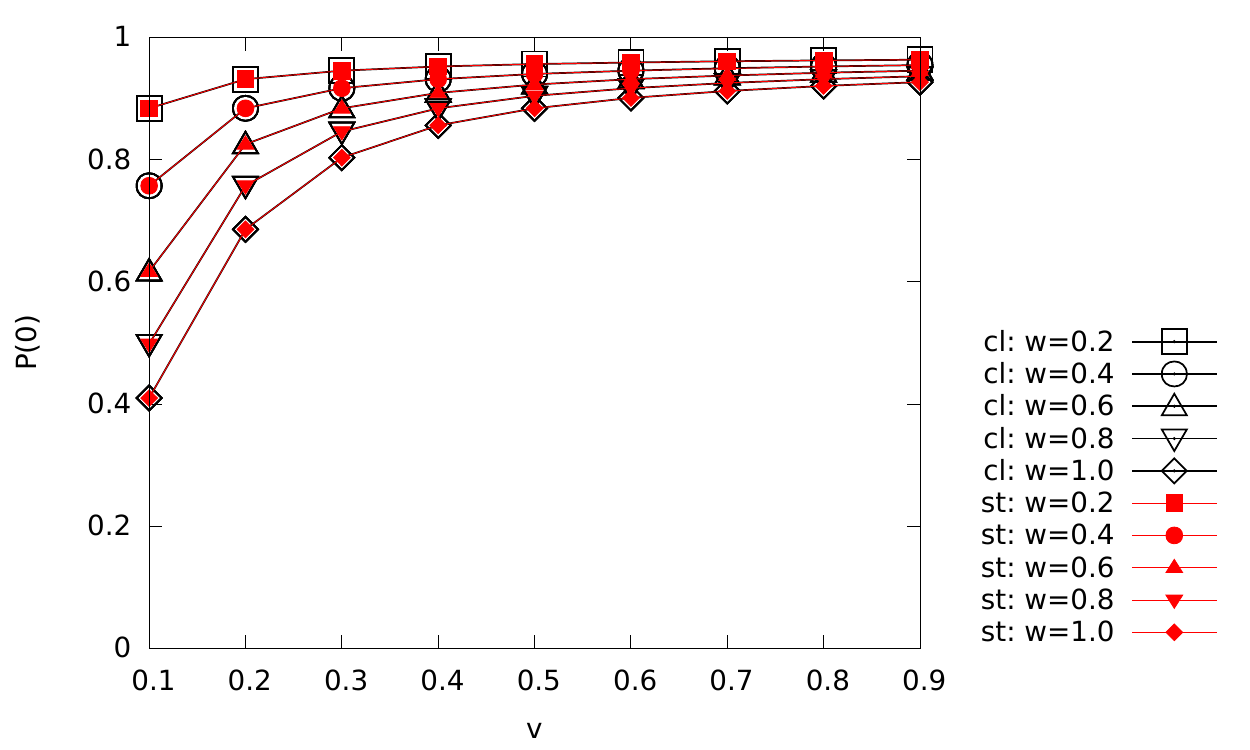}
\label{fig:7A}
}
\subfigure[$p=0.7$]{
\includegraphics[width=.5\textwidth, angle=0]{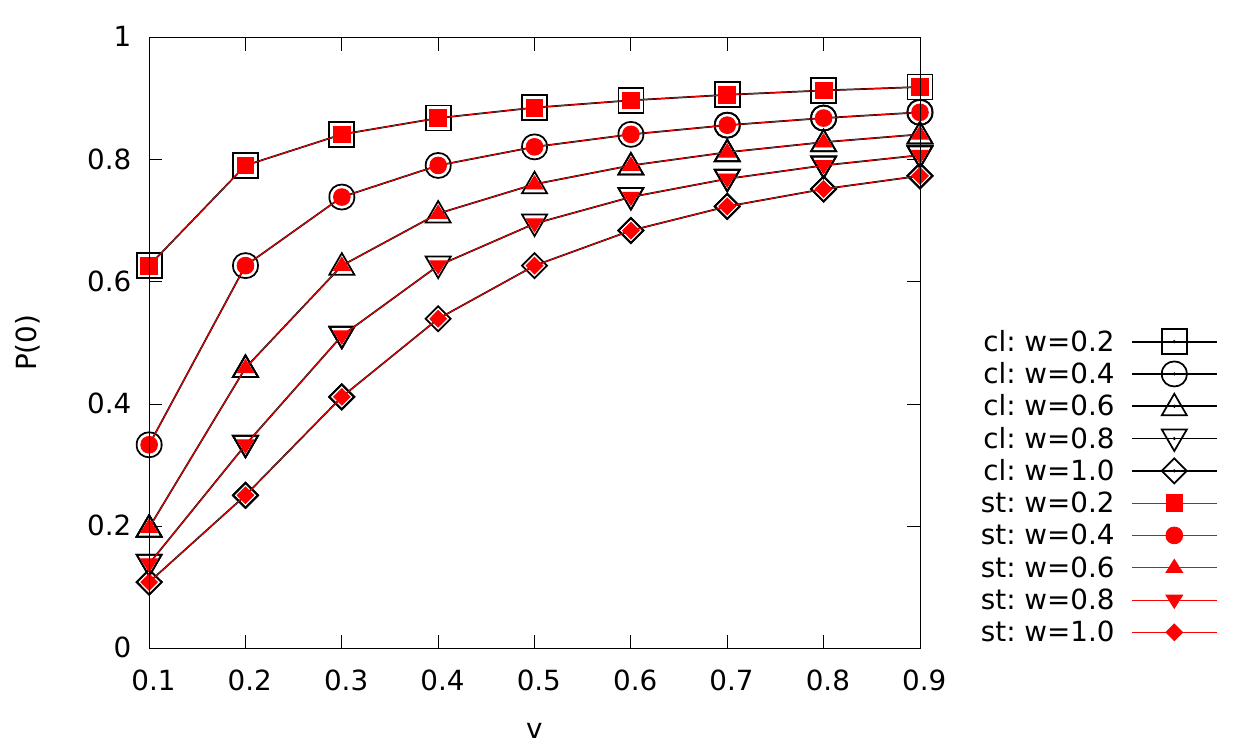}
\label{fig:7B}
}
\caption{Diagram for the real reduced network of the size $N=18$, obtained from the analysis of the state space. Here \textit{cl} refers to the network of classes, and \textit{st} to the network of states.}
\label{fig7}
\end{center}
\end{figure}

Exact calculations of $P(0)$ can be performed merely for systems much smaller than hundreds of road sections. For the sake of comparison of the methods, 
we simplified the map of Rabka leaving only nine two-way roads. This leads to the system of $2^{18}$ states. The results of the simulations for this system
are shown in \ref{fig6}. As the system is much simplified, the results for the full (Fig.\ref{fig5}) and reduced (Fig.\ref{fig6}) traffic networks
differ substantially for $p=0.1$. Surprisingly, those for $p=0.7$ are quite similar. The same simplified network is solved exactly by the solution of $2^{18}=262144$ 
Master equations \cite{vkamp} for the stationary state for different sets of 
the model parameters of the related Kripke structure. In this exact method, the parameters $p,v,w$ enter to the weights of links between states, or - equivalently -
to the rates of the processes which drive the system from one state to another. For each case we can then calculate, in accordance with Eq.\ref{e1}, the mean stationary 
probability that the road sections are passable. Obtained results are presented 
in Figs.\ref{fig7}. The same figure shows the solution for the classes of states, as described in \cite{m1,m2}. In this case, the class identification procedure allows 
for the reduction of the system size about twice, to $102400$ classes.\\

\section{Discussion}

The goal of this paper is to describe large traffic networks with a cellular automaton, where states of road sections are reduced to two: passable and jammed. The coarse-grained
character of the new automaton is close in spirit to the percolation effect. The results of our simulations allow to identify a phase transition between two macroscopic phases, 
again passable and jammed. Additionally, the calculations are repeated for a much smaller traffic network, constructed by a strong simplification of a map of a small Polish city.
These calculations are performed to compare the results with the exact solution of the stationary state, obtained by two equivalent methods. This comparison suggests, that the 
accordance of simulation with the exact solution is better for more jammed systems, i.e. more close to the phase transition. \\
\\

The drawback of our automaton is that all information about specific local conditions of traffic jams cannot be reproduced. The model captures merely the jam spreading. 
The parameters $w$ and $p$ depend on the external state, and serve as an input for the calculations. The parameter $v$ should be calibrated separately for each traffic system. 
After this calibration,  the main result of the model - the probability of the jammed phase - should be reproducible and useful to control the traffic phases. The advantage of 
the model is its simplicity, which allows to to simulate larger traffic systems in real time. \\

{\bf Acknowledgement:}
The research is partially supported within the FP7 project SOCIONICAL, No. 231288 and by the Polish Ministry of Science and Higher Education and its grants for Scientific 
Research and by PL-Grid Infrastructure.

\end{document}